# Temperature distribution measurement on three-phase contact line in liquid nitrogen using two-color temperature-sensitive paint

Shione Fujiwara[1], Yasuhiro Egami[2], Osamu Kawanami[3], Yu Matsuda[1,*]

1. Department of Modern Mechanical Engineering, Waseda University, 3-4-1 Ookubo, Shinjuku-ku, Tokyo, 169-8555, Japan
2. Department of Mechanical Engineering, Aichi Institute of Technology, 1247 Yachigusa, Yakusa-Cho, Toyota, Aichi, 470-0392, Japan
3. Department of Mechanical Engineering, University of Hyogo, 2167 Shosha, Himeji 671-2280, Japan

* Corresponding author: Yu Matsuda

**Email:** y.matsuda@waseda.jp

## Abstract

Cryogenic phase-change phenomena play an important role in a wide range of engineering applications, including cryogenic cooling systems, superconducting technologies, and space propulsion systems. In particular, the three-phase contact line is recognized as a key region governing evaporation and heat transfer. However, direct measurements of temperature distributions near cryogenic three-phase contact lines remain limited because conventional infrared thermography becomes increasingly difficult at extremely low temperatures. In this study, a two-color temperature-sensitive paint (2C-TSP) technique was applied to visualize the temperature field around a liquid-nitrogen three-phase contact line. A temperature-sensitive dye and a temperature-insensitive reference dye were incorporated into a single coating, enabling robust temperature measurements based on luminescence intensity ratios by compensating for changes in optical intensity caused by refraction and reflection at the liquid–gas interface. Temperature distributions were measured under three heating conditions with heat fluxes of $1.1 \times 10^2$, $4.3 \times 10^2$, and $9.0 \times 10^2\ \mathrm{W/m^2}$. The measured temperature fields revealed a localized temperature minimum at the observed three-phase contact line, suggesting localized cooling associated with phase change. Quantitative analysis showed that the average temperature in the liquid region remained nearly constant, whereas the temperature in the gas region increased with increasing heat flux. These observations reveal a non-uniform thermal structure around the cryogenic three-phase contact line. The present results demonstrate that 2C-TSP is a promising technique for direct visualization of temperature fields around cryogenic three-phase contact lines and provides new insights into phase-change phenomena in liquid nitrogen.

## 1. Introduction

Cryogenic fluids play an essential role in a wide range of engineering applications, including rocket propulsion using liquid hydrogen and liquid oxygen, [1] superconducting magnet systems for fusion reactors, [2] and emerging quantum technologies. [3] Accurate understanding of heat transfer and phase-change phenomena in cryogenic fluids is therefore of both fundamental and practical importance. Among cryogenic fluids, liquid nitrogen is particularly attractive because of its availability, low cost, and widespread use as a cooling medium. [4, 5] Understanding boiling and evaporation phenomena in liquid nitrogen is therefore important for the design and optimization of cryogenic cooling systems. The three-phase contact line is recognized as one of the key regions governing boiling and evaporation heat transfer. [6, 7] Previous studies have suggested that large temperature gradients and intense evaporation can occur in the immediate vicinity of the contact line. [8, 9] Although contact-line temperature distributions have been investigated for room-temperature fluids, corresponding measurements for cryogenic fluids remain limited because of the difficulty of temperature measurements at extremely low temperatures. To understand phase-change phenomena near the contact line, spatially resolved temperature measurements are required. Infrared thermography has been widely used to investigate temperature distributions in boiling and evaporation phenomena because it enables non-contact, two-dimensional temperature measurements with high spatial resolution. [8-10] However, the application of infrared thermography to cryogenic fluids remains challenging. Since thermal radiation decreases significantly with temperature, the emitted infrared intensity from cryogenic surfaces is much lower than that from room-temperature systems, resulting in reduced measurement sensitivity and increased uncertainty. Consequently, alternative temperature measurement techniques are required for cryogenic phase-change studies.

In this study, we focus temperature-sensitive paint (TSP) [11] to measure the temperature distribution at the solid–gas–liquid three-phase contact line formed in liquid nitrogen. The TSP method is a temperature measurement technique based on the temperature dependence of the luminescence intensity of fluorescent molecules and has been widely applied to temperature measurements of two-phase flows. [12-16] By selecting appropriate probe molecules, TSP is expected to be applicable to cryogenic conditions such as liquid nitrogen. In TSP measurements of two-phase flows, optical effects such as refraction and reflection at the solid–gas–liquid interface introduce errors. In our previous study, [13] these effects were corrected by applying a single correction factor. Although this approach is simple and easy to apply, the correction factor needs to be re-calibrated whenever the optical configuration changes. Therefore, in this study, two-color TSP (2C-TSP) is employed to provide a more general and robust correction for the effects of refraction and reflection. We selected a pair of dyes for 2C-TSP suitable for cryogenic environments, one exhibiting temperature-sensitivity and the other temperature-insensitive and investigated their temperature characteristics. The method was then applied to measure the temperature distribution in the liquid nitrogen meniscus formed between two parallel plates.

## 2. Experimental methods

### 2.1 Fabrication of 2C-TSP

TSP is composed of temperature-sensitive dye and a polymer binder. [11] In the TSP method, the temperature distribution on the TSP-coated surface is measured by capturing, with a camera, the temperature-dependent changes in the luminescence intensity of the dye under excitation light. In TSP measurements at solid–gas–liquid interfaces, the luminescence intensity of TSP changes independently of temperature due to reflection and refraction of both the excitation light and the emitted luminescence at the interface. Therefore, correction of these optical effects is required for accurate temperature measurements. In our previous study, we employed a correction factor to correct these effects. [13] However, this correction factor depends on the configuration of the camera and the light source and therefore changes when their arrangement is altered, limiting its general applicability. In this study, 2C-TSP was used to measure the temperature distribution at the solid–gas–liquid three-phase contact line formed in liquid nitrogen. We used a 2C-TSP consisting of a mixture of a temperature-sensitive dye and a temperature-insensitive reference dye, and determined the temperature from the ratio of their luminescence intensities. By taking the intensity ratio of the two wavelength components, this method can reduce the optical effects, such as reflection and refraction at the gas-liquid interface. Therefore, it is effective for temperature measurements in regions involving an interface. Here, we adopted Bis-(2,2,2"-terpyridine) ruthenium(II) chloride ($Ru(trpy)_3$) [11] as a temperature-sensitive dye and $BaMg_2Al_{10}O_{17}$:Eu (BAM) as a reference phosphor [17, 18] for our 2C-TSP. $Ru(trpy)_3$ has been used for temperature measurements in cryogenic wind tunnels, [19] and its sensitivity in cryogenic temperature ranges is expected to be sufficient; however, to the best of the authors' knowledge, it has not yet been applied to measurements in two-phase flows. Similarly, although BAM has been used as a reference dye in two-color pressure-sensitive paint (2C-PSP), [17] there have been no reported applications to two-color TSP or to measurements in two-phase flows. The fabricated 2C-TSP film

In this study, these dye molecules were selected for the development of 2C-TSP. We employed Clearcoat UVR (AkzoNobel, Netherlands) as a polymer binder. The 2C-TSP solution was prepared by mixing 4.2 mg of $Ru(trpy)_3$, 375 mg of BAM, 5.6 mL of ethanol (Fujifilm Wako Pure Chemical Corporation, Japan), 1.0 mL of Hardener S 66/22R (AkzoNobel, Netherlands), 2.0 mL of Clearcoat UVR (AkzoNobel, Netherlands), and 0.6 mL of Thinner C 25/90S (AkzoNobel, Netherlands). This 2C-TSP solution was spin-coated onto a glass substrate at a maximum rotational speed of 1000 rpm. Then, the surface properties of the fabricated 2C-TSP film, including film thickness, surface roughness, and surface morphology observed by field emission scanning electron microscopy (FE-SEM; JSM-7001F, JEOL Ltd., Japan) were characterized. The thickness of the fabricated film was measured to be $5.1\ \mu\mathrm{m}$ using a profilometer (ET-200, Kosaka Laboratory, Japan). The surface roughness was also measured as an arithmetic mean roughness (Ra) of $0.04 \pm 0.01\ \mu\mathrm{m}$ and a maximum height roughness (Rz) of $0.20 \pm 0.07\ \mu\mathrm{m}$, where the errors represent the standard deviations. The low surface roughness is also apparent in the SEM image of the film surface shown in Fig. 1.

Since the wettability of the surface affects the meniscus shape and the position of the three-phase contact line, the contact angle of liquid nitrogen on the fabricated 2C-TSP surface was measured. Here, the contact angle was estimated using the capillary rise method. When two parallel plates with a gap distance of 0.15 mm were inserted into liquid nitrogen, the capillary rise height was approximately 10 mm. Using the liquid nitrogen density of $808\,\mathrm{kg/m^3}$ and the surface tension of $8.96\,\mathrm{mN/m}$, [20] the contact angle was calculated to be 49°.

## 2.2 Calibration curve of the fabricated 2C-TSP

The emission spectrum of the fabricated 2C-TSP film was measured using a fluorescence spectrometer (Quantaurus-Tau, C11367, Hamamatsu Photonics, Japan). The temperature of the 2C-TSP film was controlled using a cryostat (CoolSpeK, UPS-203B, Unisoku, Japan). The measured emission spectrum is shown in Fig. 2. As shown in the figure, the emission peak of $Ru(trpy)_3$ was observed around 610 nm, while an emission peak of BAM was observed around 450 nm, indicating a clear spectral separation between the two components. When the temperature was varied from −150 to −195 °C, the luminescence intensity of $Ru(trpy)_3$ increased with decreasing temperature, whereas that of BAM showed little change. These results confirmed that the two-color intensity ratio method using $Ru(trpy)_3$ as the temperature-sensitive dye and BAM as the reference dye is appropriate for cryogenic measurements. In the two-color intensity ratio method, the measured temperature can be corrected using the ratio-of-ratios approach. [21] The relation between the ratio-of-ratios and temperature is written as

$$\frac{R_{\mathrm{ref}}}{R} = a\left(\frac{T}{T_{\mathrm{ref}}}\right)^{b} + (1-a), \tag{1}$$

where $R$ is defined as the ratio of the luminescence intensity of $Ru(trpy)_3$, $I_{\mathrm{Ru}}$, to that of BAM, $I_{\mathrm{BAM}}$ ($R = I_{\mathrm{Ru}}/I_{\mathrm{BAM}}$). The subscript "ref" indicates a reference condition and $T$ is the temperature. The temperature at the reference condition was $-165$°C in this study. The experimental data and the fitted calibration curve are shown in Fig. 3. The shaded region denotes the 95% prediction interval, which includes both the uncertainty of the fitted calibration curve and the scatter of the experimental data. The constants $a$ and $b$ were determined to be $0.21 \pm 0.07$ and $3.3 \pm 0.7$, respectively, where the uncertainties represent the 95% confidence intervals. The uncertainty in the temperature measurement, estimated from the 95% prediction interval of the calibration model, was approximately 2.5%.

## 2.3 Experimental setup

The experimental setup is shown in Fig. 4. In this study, we measured the temperature distribution in the liquid nitrogen meniscus formed between two parallel glass plates. The dimensions of the glass plates were 75.3 mm in length, 25.3 mm in width, and 1.05 mm in thickness. In this study, a reservoir was filled with liquid nitrogen, and the two plates were placed such that their lower ends were immersed in the liquid nitrogen. The gap between the two plates was fixed at 0.15 mm using a shim plate. The two plates were installed with a tilt angle of 13° from the vertical direction to facilitate visualization of the meniscus. In addition, the meniscus position was maintained below the opening of the container so that the interior of the container was filled mostly with nitrogen gas, thereby preventing condensation of water on the glass surface. One of the two glass plates was coated with an indium tin oxide (ITO) film on the outer surface of the plate to facilitate electrical connections, while a 2C-TSP film was applied to the opposite surface facing the gap between the two plates. The electrical resistance of the ITO film, measured using a digital multimeter (DM2571, NF Corporation, Japan), was $130\ \Omega$. The 2C-TSP film surface was brought into contact with the meniscus, and the temperature distribution at the meniscus was measured. The TSP film was illuminated by an LED light source (IL-106UV4, HARDsoft Microprocessor Systems, Poland). The luminescence images of the temperature-sensitive and reference dyes emitted from the 2C-TSP film were simultaneously acquired in the same camera frame using a CCD camera (BU-52LN, BITRAN, Japan) through a stereoscope-like device consisting of four mirrors and two optical filters. The details of the stereoscope-like device were described previously. [13] The stereoscope-like device used in the present study enables two images of an identical area to be obtained at different wavelength bands. Specifically, one image was the luminescence image of Ru(trpy)3 detected through an optical band-pass filter (transmissive wavelength: 585 ± 32.5 nm, ET585/65m, Chroma Technology, USA), whereas the other was the luminescence image of BAM detected through an optical band-pass filter (transmissive wavelength: 460 ± 25 nm, AT460/50m, Chroma Technology, USA). The two wavelength components were acquired simultaneously from the same viewing direction, thereby suppressing the influence of light reflection and refraction near the contact line caused by changes in the interface shape. In the present optical system, images were acquired using the CCD camera. The images were recorded with an exposure time of 0.3 s and the spatial resolution was 86 μm/px.

## 3. Results and Discussion

The measured temperature distributions are shown in Fig. 5, where the heat fluxes generated by the ITO film were $1.1 \times 10^2$, $4.3 \times 10^2$, and $9.0 \times 10^2\ \mathrm{W/m^2}$. As shown in the figures, a localized temperature decrease can be clearly observed in the vicinity of the three-phase contact line. The location of the temperature minimum coincided with the observed three-phase contact line, indicating localized cooling associated with phase change near the three-phase contact line. To quantitatively evaluate the temperature field, representative regions in the gas and liquid regions were selected, and the average temperatures within these regions were calculated. The average temperature in the liquid phase was approximately $-192$ ℃ for all heating conditions. Although this value is slightly higher than the saturation temperature of liquid nitrogen at atmospheric pressure, the difference is considered reasonable when taking into account the uncertainty of the calibration curve and the signal-to-noise ratio of the measured data. The standard deviation of the measured temperature in the liquid region was approximately $4$ ℃, which is comparable to the deviation from the saturation temperature. Therefore, the observed discrepancy is considered to be within the uncertainty of the present measurement system. Fig. 6 shows the temperature difference between the gas and liquid regions as a function of the imposed heat flux. As the input heat flux increased, the average temperature in the gas region increased, whereas the temperature in the liquid region remained nearly constant. As a result, the mean temperature difference between the gas and liquid regions increased monotonically with increasing imposed heat flux. Although the error bars, calculated as the standard deviation of the temperature distribution, are relatively large because of measurement fluctuations, the observed trend suggests that the temperature in the gas region was more sensitive to the applied heating than that in the liquid region, which remained close to the saturation temperature of liquid nitrogen.

The contact-line temperature was also evaluated from the manually selected contact-line region. To avoid excessive sensitivity to measurement noise, the contact-line temperature was defined as the average of the lowest 20% of the temperature values within the selected region. The resulting temperature was approximately $-195^\circ\mathrm{C}$, which was slightly lower than the average temperature in the liquid region. However, the standard deviation within the contact-line region was relatively large, approximately $8$ ℃. Therefore, although the result supports the presence of a local temperature depression near the contact line, the quantitative value of the contact-line temperature should be interpreted with caution.

## Conclusions

A two-color temperature-sensitive paint (2C-TSP) technique was applied to visualize the temperature field around a liquid-nitrogen three-phase contact line. The temperature distributions under three heating conditions were successfully measured, demonstrating the applicability of the 2C-TSP method to cryogenic phase-change phenomena.

A localized temperature decrease was consistently observed in the vicinity of the three-phase contact line, and the location of the temperature minimum coincided with the observed contact-line position. This result suggests localized cooling associated with phase change near the contact line. Quantitative analysis of representative gas and liquid regions showed that the liquid-region temperature remained nearly constant, whereas the gas-region temperature increased with increasing heat flux. The contact-line temperature was slightly lower than the average liquid-region temperature. However, the uncertainty associated with the contact-line temperature remained relatively large because of measurement noise.

The present results demonstrate that 2C-TSP is a promising tool for direct visualization of temperature fields near cryogenic three-phase contact lines. Although the technique provided valuable temperature-field information, quantitative estimation of local heat-flux distributions remains challenging because the temperature difference across the glass substrate is much smaller than the measurement uncertainty. Further improvements in temperature resolution and thermal analysis methods will be necessary for reliable heat-flux measurements.

## Acknowledgement

This work partially supported by the JSPS KAKENHI Grant Numbers 23K03679. The experiment was supported by Ryota Asami and Ryo Ozaki.

## Declaration of competing interest

The authors declare that they have no known competing financial interests or personal relationships that could have appeared to influence the work reported in this paper.

## Figures and Tables

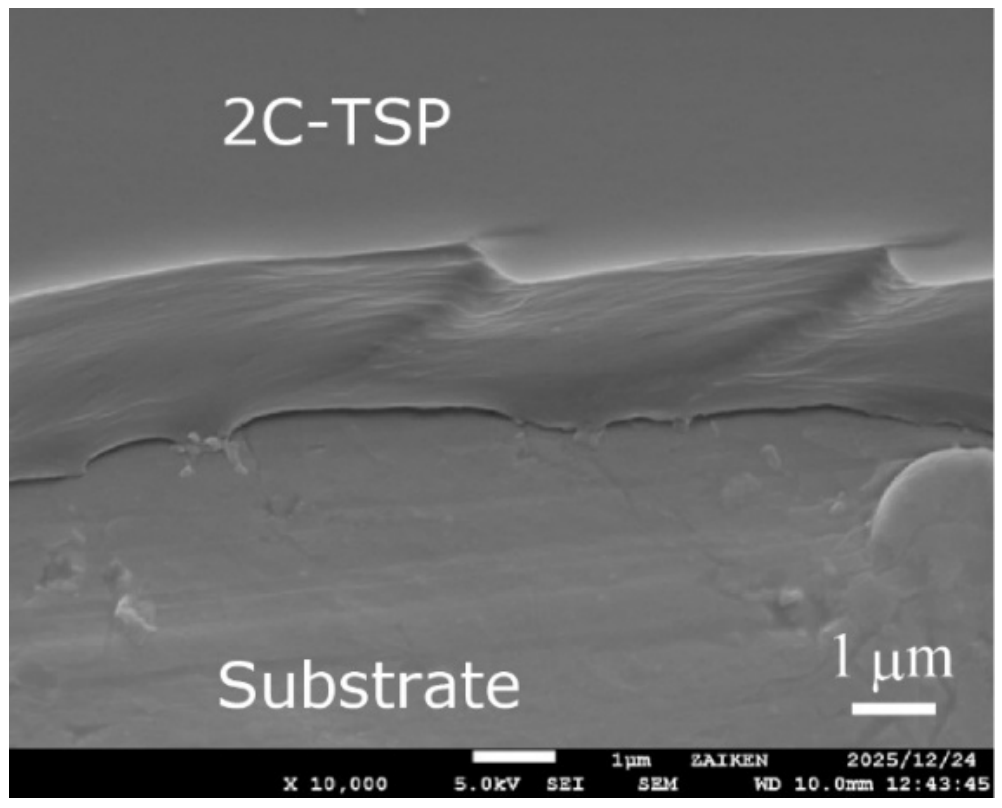

**Figure 1.** Representative FE-SEM image of the fabricated 2C-TSP film.

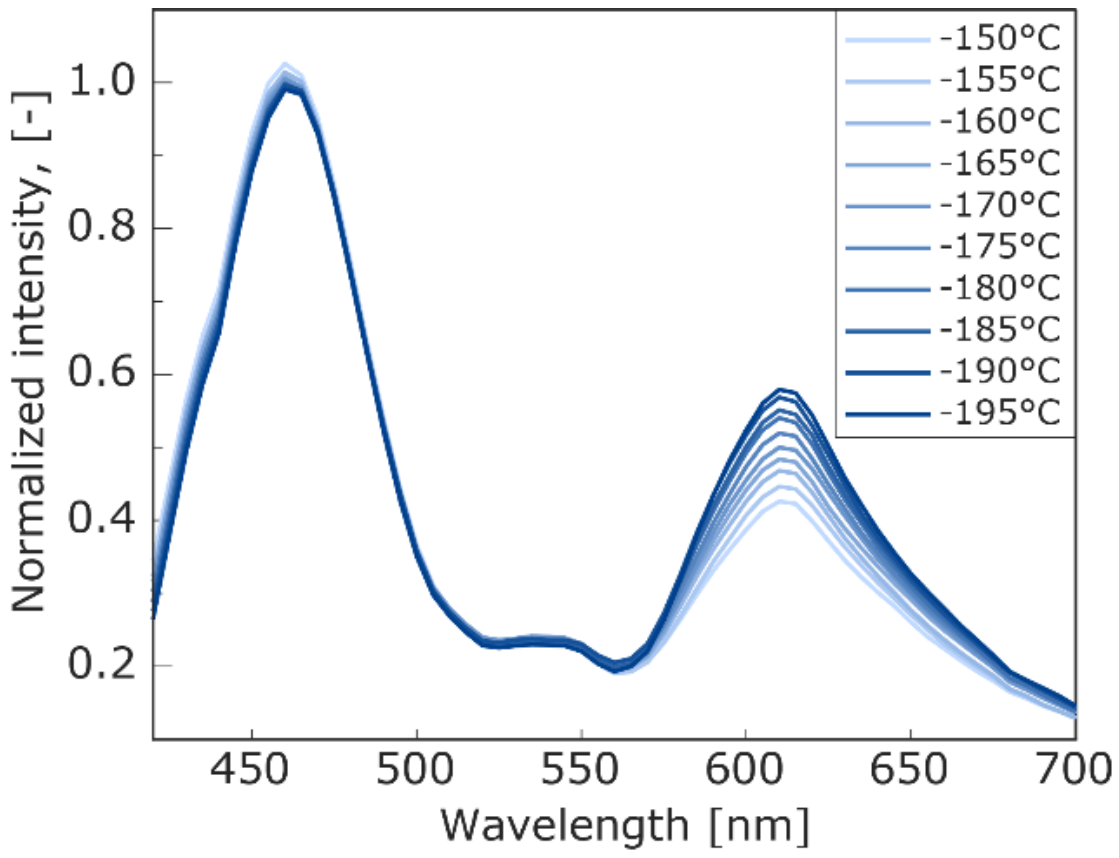

**Figure 2.** Emission spectrum of the fabricated 2C-TSP film.

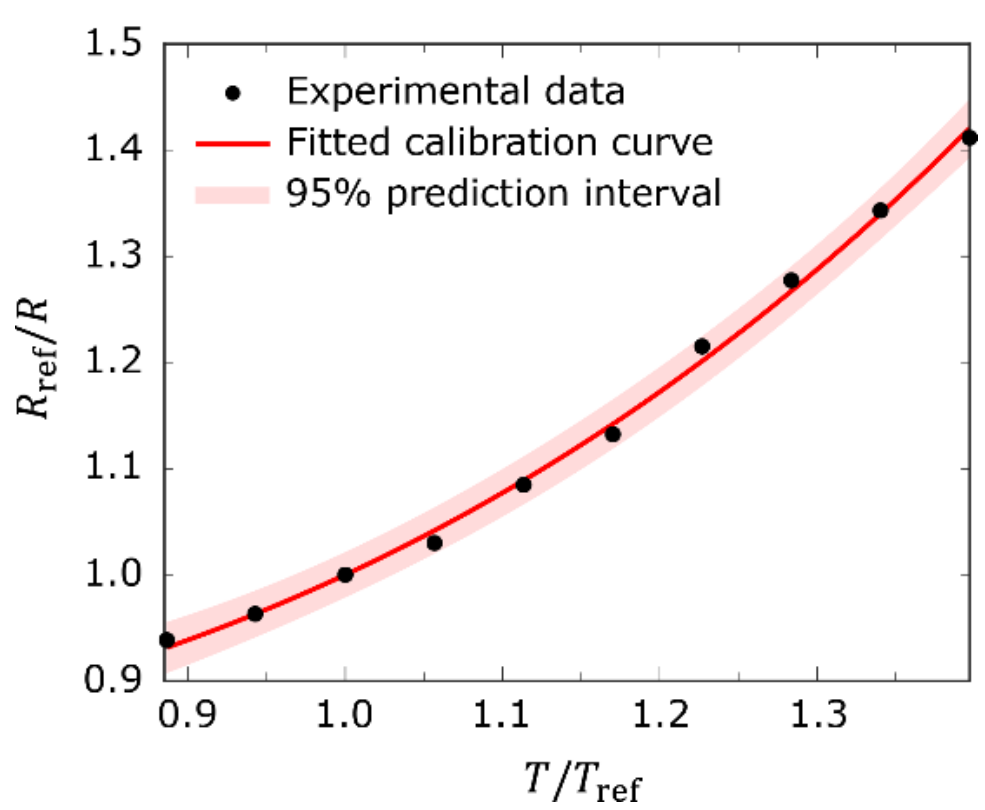

**Figure 3.** Calibration curve of the 2C-TSP film

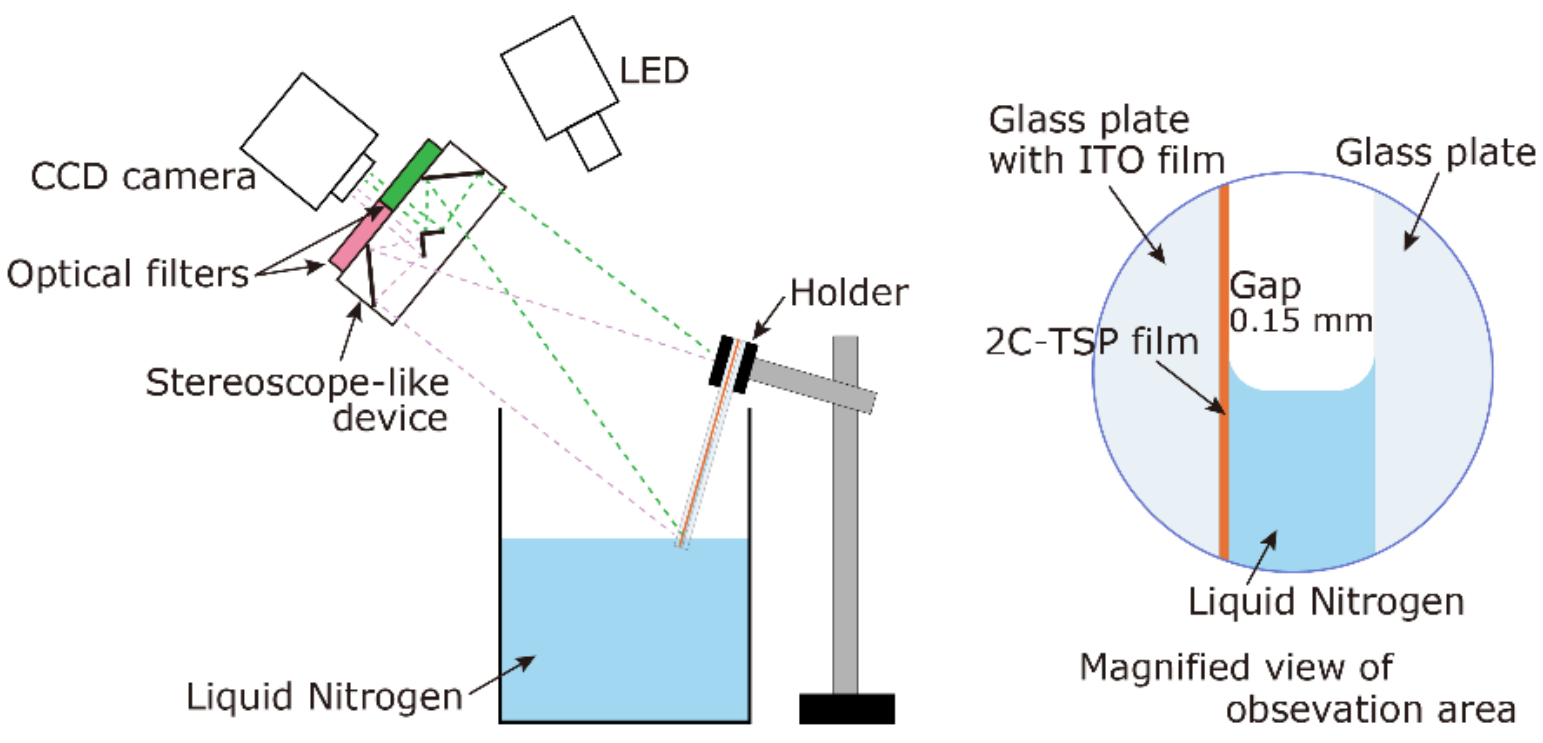

**Figure 4.** Schematic illustration of the experimental setup.

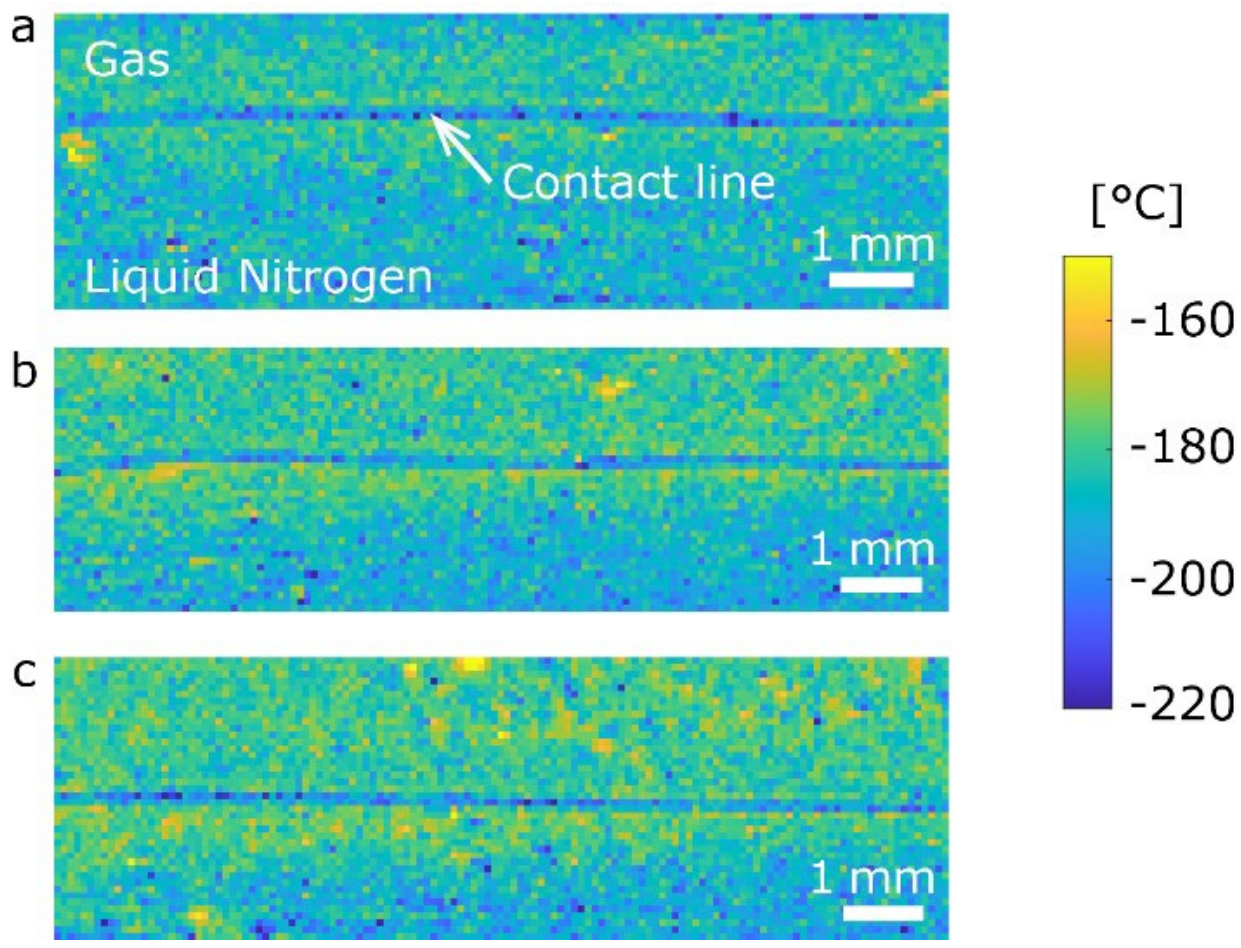

**Figure 5.** Results of temperature measurement using 2C-TSP. The input heat fluxes were (a) 110 W/m$^2$, (b) 430 W/m$^2$, and (c) 900 W/m$^2$.

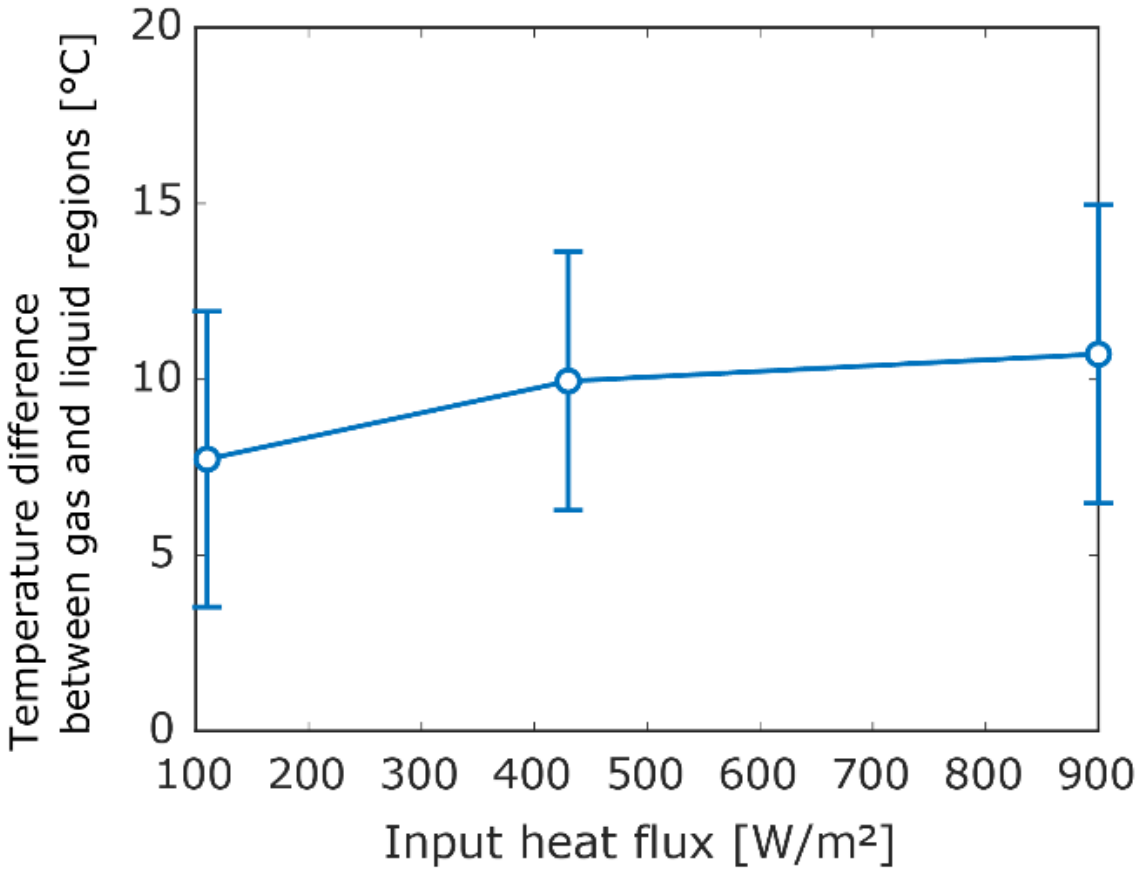

**Figure 6.** Temperature difference between the gas and liquid regions versus imposed heat flux.